\documentclass[manuscript]{aastex}
\usepackage{amsmath,amssymb,amstext}

\usepackage[breaklinks,colorlinks,citecolor=blue,linkcolor=magenta]{hyperref} 

\usepackage[all]{hypcap} 

\usepackage{aas_macros}
\usepackage{natbib}
\shorttitle{Investigating 2MASS J06593158-0405277: a FUor burst in a triple system?}
\shortauthors{Caratti o Garatti et al.}

\begin{document}

\title{Investigating 2MASS J06593158-0405277: a FUor burst in a triple system?}

\author{A. Caratti o Garatti\altaffilmark{1}, R. Garcia Lopez\altaffilmark{1}, T.P. Ray\altaffilmark{1}, J. Eisl\"{o}ffel\altaffilmark{2}, B. Stecklum\altaffilmark{2}, A. Scholz\altaffilmark{3},
S. Kraus\altaffilmark{4}, G. Weigelt\altaffilmark{5}, A. Kreplin\altaffilmark{4}, V. Shenavrin\altaffilmark{6}}
\affil{\altaffilmark{1}School of Cosmoc Physics, Dublin Institute for Advanced Studies, 31 Fitzwilliam Place, Dublin 2, Ireland}
\affil{\altaffilmark{2}Th\"{u}ringer Landerssternwarte Tautenburg, Sternwarte 5, Tautenburg, Germany}
\affil{\altaffilmark{3}University of St Andrews, School of Physics and Astronomy, North Haugh, St Andrews KY16 9SS, UK}
\affil{\altaffilmark{4}School of Physics, University of Exeter,Physics Building, Stocker Road, Exeter, EX4 4QL, UK }
\affil{\altaffilmark{5}Max Planck Institut f\"{u}r Radioastronomie, Auf dem H\"{u}gel 69, 53121 Bonn, Germany}
\affil{\altaffilmark{6}Lomonosov Moscow State Univ., Sternberg Astron. Inst., Universitetsky pr. 13, 119234 Moscow, Russia}
\email{alessio@cp.dias.ie}

\begin{abstract}
FUor outbursts in young stellar objects (YSOs) are the most dramatic events among episodic accretion phenomena. 
The origin of these bursts is not clear: disk instabilities and/or disk perturbations by an external body being the most viable hypotheses. 
Here, we report our VLT/SINFONI high angular resolution AO-assisted observations
of 2MASS\,J06593158-0405277, which is undergoing a recently discovered FUor outburst.
Our observations reveal the presence of an extended disc-like structure around the FUor, a very low-mass companion (2MASS\,J06593158-0405277B) at $\sim$100\,au in projection, and, possibly, a third closer companion at $\sim$11\,au. 
These sources appear to be young, displaying accretion signatures.
Assuming the components are physically linked, 2MASS\,J06593158-0405277 would then be one of the very few triple systems observed in FUors.

\end{abstract}


\maketitle
\section{INTRODUCTION}

FU Orionis-like outbursts (hereafter FUors) are the most dramatic events among episodic accretion phenomena in YSOs, displaying a mass accretion rate ($\dot{M}_{acc}$) increase by several orders 
of magnitude~\citep[up to $\dot{M}_{acc}\sim$10$^{-4}$\,M$_\odot$\,yr$^{-1}$; see e.g.][]{H-K,audard}. As a result,
the YSO brightness is boosted by 5 magnitudes or more in the optical~\citep[e.g.,][]{herbig77}.
These bursts have a dramatic impact on the YSO circumstellar environment
~\citep[][]{baraffe,vorobyov13,audard}.
The rise times of these outbursts are usually short (from several months to a few years), whereas the decay timescales range from years to several decades. 
Spectra of FUor progenitors have only been obtained in two cases, both
indicating that they are low-mass YSOs~\citep[e.g.,][]{herbig77,miller}.

The origin of bursts in FUors is not well understood, and
disk instabilities, disk fragmentation, and/or disk perturbation by an external body are the most viable hypotheses~\citep[see review by][and references therein]{audard}. 
Several models with thermal or gravitational plus magneto-rotational instabilities are able to mimic the main characteristics of the outbursts~\citep[see, e.g.,][]{clarke,zhu10b,zhu10a}. Notably, the recent version of the disk gravitational instability
and fragmentation model by \citet{vorobyov15} reproduces the physical properties including the episodic nature of the bursts. Aside from internal instabilities, external triggers, namely
close encounters in binary or multiple systems, can produce FUor bursts~\citep[][]{bonnell92,reipurth04,reipurth14}.

FUors are rare transient phenomena, and they have usually been detected and studied in detail several months or even years after the beginning of the outburst. 
\objectname{2MASS J06593158-0405277} is a YSO undergoing an FU Ori-type outburst~\citep{maehara,hillenbrand14}. The outburst was detected on the 23rd of November 2014~\citep{maehara}, and
previous observations indicate that the object began its slow rise at the end of 2013 and reached its maximum at the end of 2014~\citep{hackstein,kospal15}.
From archival photometric data, \citet{kospal15} identify the progenitor as a low-mass T Tauri star (0.75\,M$_\odot$) with a circumstellar disk of 0.01--0.06\,M$_\odot$, and an age of $\sim 6 \times$10$^5$\,yr.

\section{OBSERVATIONS AND DATA REDUCTION}
\label{observations}

\objectname{2MASS J06593158-0405277} was observed with ESO/VLT UT4 on 2015 January 25 with
the near-infrared integral field spectrograph SINFONI~\citep[][]{sinfoni} in the $J$ (1.1--1.4\,\micron) and $K$ (1.95--2.45\,\micron) bands with a spectral resolution of $ \mathrm R\sim 2000$ and 4000, respectively. The observations were performed with the highest spatial sampling (12.5 milliarcseconds - mas - pixel scale, and field of view - FoV - of 0\farcs8$\times$0\farcs8) and in AO-assisted mode (using the target as a natural guide star) with a mean optical seeing of $\sim 0\farcs8$, which provided a full-width at half maximum (measured on the point spread function - PSF) of $\sim 50$\,mas and $\sim 70$\,mas in the $J$ and $K$ band, respectively. 
The detector integration time (DIT), number of sub-exposures per frame (NDITS) and exposures were 30\,s, 4,
and 5 ($J$ band), and 15\,s, 5, and 7 ($K$ band), for a total 
integration time on-source of 600\,s and 525\,s in the $J$ and $K$ band, respectively.
An additional photometric standard star (HD 42655) was observed for flux calibration of the data cubes and removal of telluric features from the spectra.
Data were reduced using the SINFONI data reduction pipeline in GASGANO~\citep{modigliani}, that
is, dark and bad pixel masks, flat-field corrections, optical distortion correction, and wavelength calibration with arc lamps.

Two spectra were extracted from each data cube using a 5$\times$5 pixel area centered at the position
of the peak flux of the FUor and its companion ($B$, see Sect.~\ref{results}). 
Due to the FUor brightness and its spatially extended disk-like emission (see Sect.~\ref{results}), a background spectrum was extracted close to the companion position and subtracted from its spectrum to remove the contamination from the FUor source.

Additionally, continuum images (at 1.25 and 2.2\,$\mu$m) and continuum-subtracted Pa$\beta$ and
Br$\gamma$ images were created following the same procedures as in \citet{beck08} and \citet{rebeca13}.

As no PSF standard was observed, to remove the FUor's PSF from the continuum and evaluate the asymmetry of the bright central core (see Sect.~\ref{results}), we rotate the continuum images by 180$\degr$, match the PSF and subtract the rotated images from the original ones in both $J$ and $K$ bands.

\section{RESULTS}
\label{results}

\begin{figure*}
\includegraphics[width=17cm]{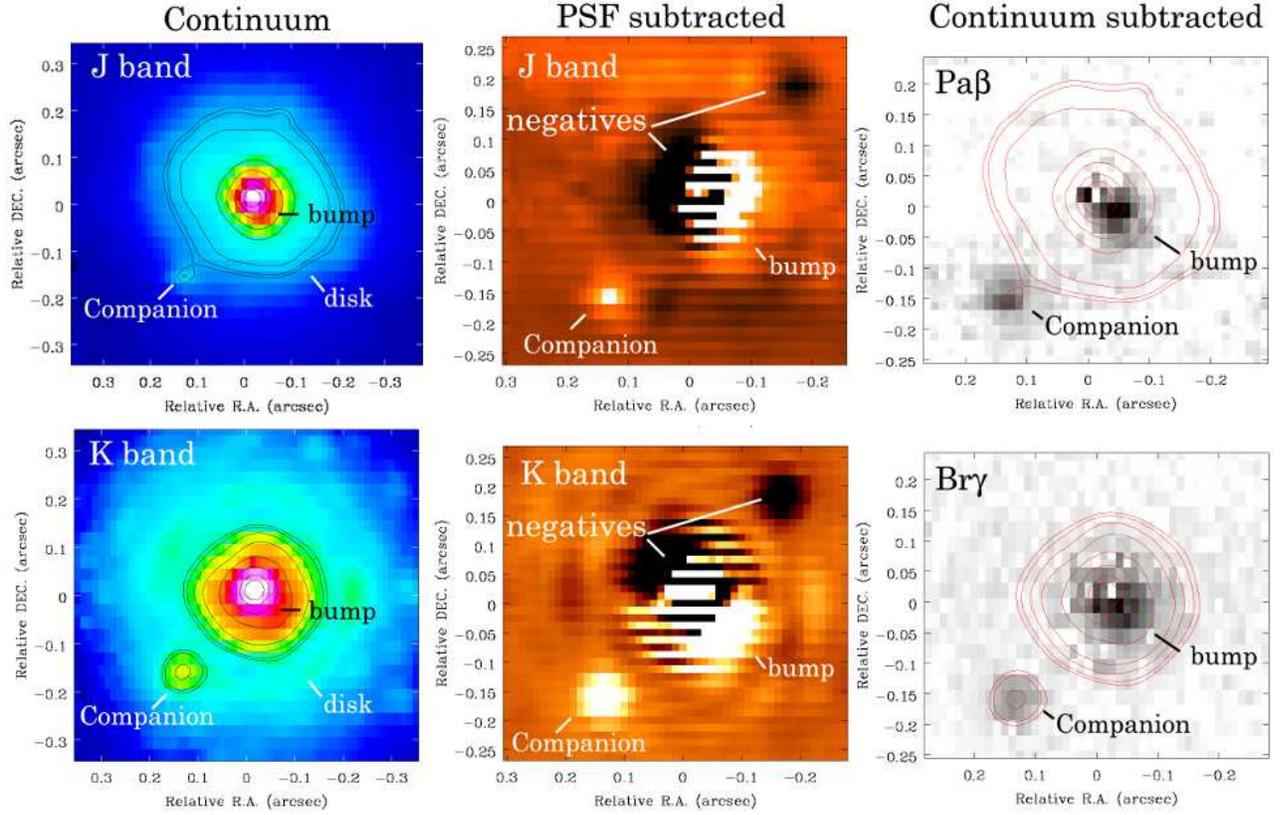}
\caption{\textit{Left}: SINFONI $J$- (upper panel) and $K$-band (lower panel) images of 2MASS\,J06593158-0405277 showing the disk-like feature, the ``bump'' position, and the companion ($B$). \textit{Center}: Difference $J$- (upper panel) and $K$-band (lower panel) images showing the `bump'. 
The PSF was removed by subtracting the same image rotated by 180$\degr$.
\textit{Right}: Pa$\beta$ (upper panel) and Br$\gamma$ (lower panel) continuum-subtracted gray-scale images. The same continuum contours are shown as in the left panels.\label{fig1}}
\end{figure*}

As illustrated in Fig.~1, SINFONI data cubes in both bands show three main features around 2MASS\,J06593158-0405277: \textit{i)} a faint companion (hereafter 2MASS\,J06593158-0405277B or $B$) 
located at 227\,mas~\citep[or 102\,au in projection at a distance of 450\,pc~\footnote{The distance to the object is highly uncertain as well as its association with the molecular cloud \objectname{L 1650},
d$\sim$2.3\,Kpc~\citep[][]{kim}}, see][]{kospal15}, 
southeast from the central source (see left panels of Fig.~\ref{fig1}); \textit{ii)} an extended and elliptical disk-like structure around the FUor, 
(size $\sim$0\farcs2 or 90\,au at a distance of 450\,pc), possibly scattered light from the disk or circumstellar nebulosity~\citep[e.g.][see also left panels of Fig.~1]{goodrich}; \textit{iii)} a bright asymmetric central core of the FUor source, which has a `bump' towards the southwest, clearly detected after subtracting the FUor's rotated images (see central panels of Fig.~1). 
Moreover, both Pa$\beta$ and Br$\gamma$ lines in emission are detected at the position of 2MASS\,J06593158-0405277B  
and near the continuum emission of the `bump', as shown in the right panels of Fig.~1.



\subsection{The Companion: 2MASS J06593158-0405277B}

\begin{figure*}
\includegraphics[width=17cm]{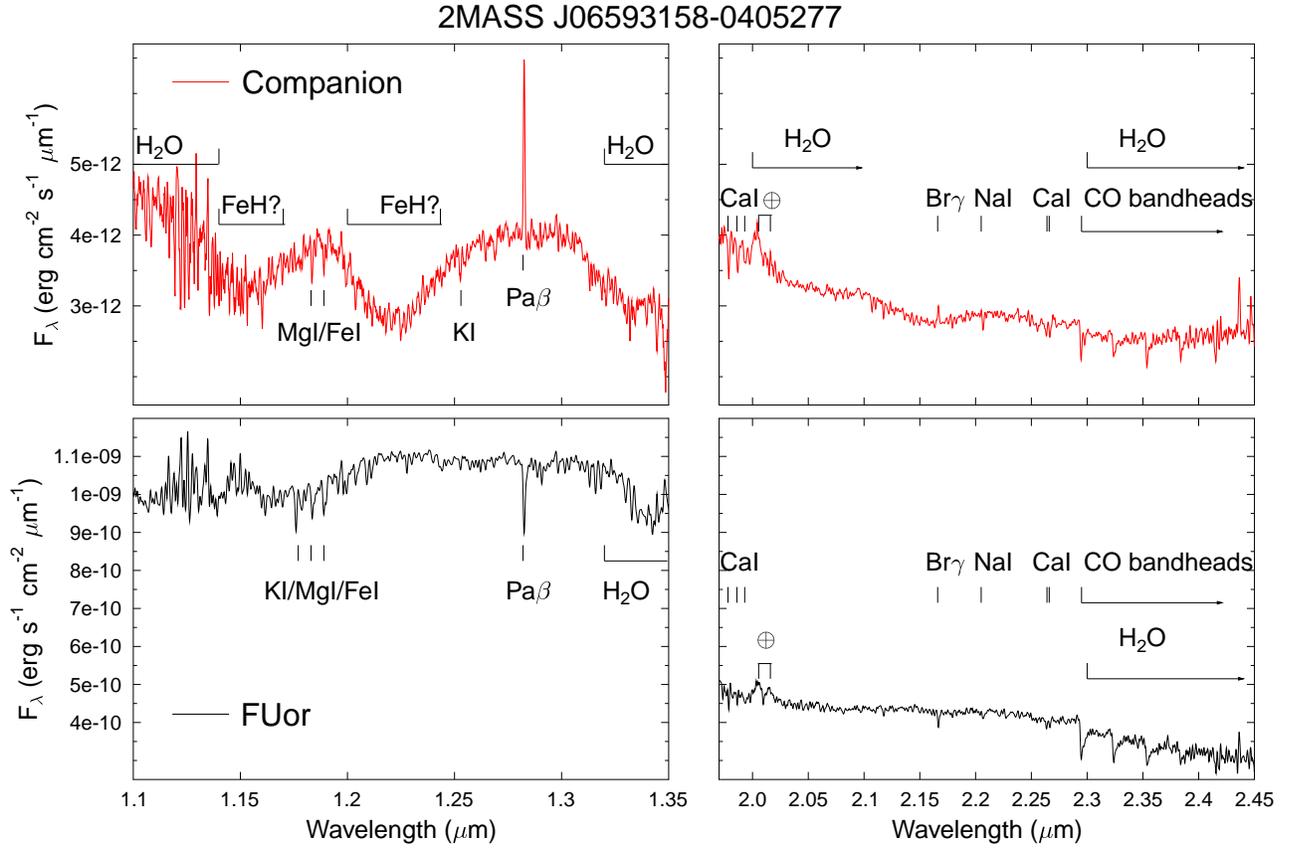}
\caption{SINFONI $J$- and $K$-band flux calibrated spectra of 2MASS\,J06593158-0405277B (red solid line, upper panels) and of the FUor (black solid line, lower panels). The most prominent features are also labeled.\label{fig2}}
\end{figure*}

On the basis of Gaussian fitting of both $J$ and $K$-band data cubes, we infer
the following coordinates for $B$: R.A.(J2000)=06:59:31.599, Dec.(J2000)=-04:05:28.01, that is, its separation from the FUor is $\Delta \alpha$=0\farcs15, $\Delta \delta$=-0\farcs17, with a P.A.=131\fdg4.
Our PSF fitting allows us to derive $J$ and $K$-band magnitudes of 14.8$\pm$0.4\,mag and 12.9$\pm$0.2\,mag, respectively. The photometric uncertainties mostly arise from the bright background due to
the scattered light from the FUor.
In the upper panel of Figure~\ref{fig2}, we show the flux calibrated spectrum in the $J$ and $K$ bands. The spectrum was obtained after subtracting the FUor scattered light in the background (see Sect.~\ref{observations}).  
The spectrum shows several broad molecular absorption bands from H$_2$O (1.1, 1.4, 1.9, and 2.4\,$\mu$m), CO (bandheads v=2-0, 3-1, 4-2, 5-3 from 2.29\,$\mu$m in the $K$ band), and, possibly, FeH in the $J$ band,
as well as atomic absorption lines (\ion{Ca}{1} and \ion{Na}{1}) in the $K$ band. 
These absorption features have a photospheric origin, and are distinctive of late M-type stars~\citep[see e.g.,][]{cushing,rayner09,scholz09}. We use near-IR spectra of very low-mass and brown-dwarfs YSOs from \citet{scholz12} to constrain the spectral type of $B$.
The inferred spectral type (SpT) is between M5 and M9, that is SpT=M7$\pm$2, 
with $T_{\rm e}$$\sim$2900$\pm$300\,K~\citep[][]{lumhan}. To get a rough estimate of the stellar parameters, we assume that
$B$ is coeval to the FUor, thus with an age between 5$\times$10$^{5}$ and
10$^{6}$\,yr~\citep[][]{kospal15}. We adopt the evolutionary tracks from \citet{baraffe15}, inferring $M_* = 0.1 \pm^{0.1}_{0.08}$\,M$_\sun$ and $R_* = 1.0 \pm^{0.7}_{0.6}$\,R$_\sun$.


We use the $J-K$ color excess of 2MASS\,J06593158-0405277B and an SpT=M7~\citep[with $(J-K)_0$=1; see e.g.,][]{scholz12} to estimate both upper and lower values of the visual extinction ($A_{\rm V}$) towards the source, adopting the extinction law from \citet{rieke85}. Indeed, by assuming that the observed $E(J-K)$ is only due to the visual extinction, we get an upper limit of $A_{\rm V}$=5\,mag from $A_{\rm V} = [(J-K) - (J-K)_0] / 0.17$. This is only an upper limit, because this source is young and thus part of its IR excess originates from veiling ($r_\lambda$ = $F_{ex,\lambda}/F_{*,\lambda}$). The IR excess is thus:

\begin{equation}
\label{eq1}
E(J-K) = (J-K) - (J-K)_0 + 2.5 log \left(\frac{1+r_J}{1 +r_K} \right)
\end{equation}

We cannot derive both $r_J$ and $r_K$ values. However, by assuming $r_J$=0 and fitting an SpT=M7 template, we then get $r_K$$\sim$2.6, and we can infer a lower limit for the visual extinction of $A_{\rm V}$=2.4\,mag.     

The spectrum of 2MASS\,J06593158-0405277B (see upper panel of Fig.~\ref{fig2}) also displays \ion{H}{1} lines in emission, namely the Pa$\beta$ line at 1.282\,$\mu$m ($F$ = (2.4$\pm$0.1)$\times$10$^{-15}$\,erg\,s$^{-1}$\,cm$^{-2}$) and the Br$\gamma$ line at 2.166\,$\mu$m ($F$ = (6$\pm$2)$\times$10$^{-15}$\,erg\,s$^{-1}$\,cm$^{-2}$).     
These lines are associated with ongoing accretion~\citep[see, e.g.,][]{muzerolle98,natta04,caratti12}, and can be employed to derive an estimate of the mass accretion rate~\citep[see e.g.,][]{natta04,natta06}.
With the empirical relations by \citet{muzerolle98}, this translates into an average accretion
luminosity of $L_{acc}$=0.005-0.01\,L$_\sun$. The mass accretion rate can then be expressed as 
$\dot{M}_{acc}=1.25 \times L_{acc}R_*/GM_*$~\citep[see e.g.,][]{caratti12}. Inferred values 
are between 4$\times$10$^{-9}$ and 8$\times$10$^{-10}$\,M$_\sun$\,yr$^{-1}$,
consistent with accretion rates found in young brown dwarfs~\citep[][]{natta04,mohanty05,joergens}. 
  
\subsection{The Disk, the Bump, and the FUor}

$J$ and $K$ band continuum images in Fig.~\ref{fig1} (left panels) display an elongated disk-like
structure extending up to $\sim$200\,mas (or $\sim$90\,au) from the central source. 
Elliptical isophotes were fitted to the disk-like feature using the iraf STSDAS task \textit{isophote},
starting from radii larger than the central PSF (i.e. $\sim$50\,mas and $\sim$70\,mas in the $J$- and $K$-band, respectively). The isophotal fit in the $J$-band provides us with a semi-major axis P.A. of 52$\degr$$\pm$3$\degr$, 
and a ratio of the minor to major axes of 0.88$\pm$0.05. Assuming circular symmetry, the system axis inclination with respect to the line-of-sight is i=28$\degr$$\pm$6$\degr$. Results of the $K$ band analysis are less accurate but consistent with the previous ones, yielding P.A.=48$\degr$$\pm$10$\degr$ and i=22$\degr$$\pm$12$\degr$. Our analysis thus indicates that this disk-like structure is almost face-on.  

Another notable circumstellar feature around 2MASS\,J06593158-0405277 is the `bump' detected in continuum and line emission images (see Fig.~\ref{fig1}, central and right panels). This feature is not spatially resolved in the $J$ and $K$ band images, but it is clearly detected after subtracting the rotated images
(Fig.~\ref{fig1}, central panels) as an arc-shaped elongated feature, peaking at P.A$\sim$233$\degr$ and $\sim$225$\degr$ in the $J$ and $K$ bands, respectively, and with m$_J$$\sim$12.9\,mag and m$_K$$\sim$10.6\,mag.
In the continuum-subtracted Pa$\beta$ and Br$\gamma$ images (Fig.~\ref{fig1}, right panels), the `bump' emission appears as a point-like source, well fitted with a Gaussian profile. Its coordinates are R.A.(J2000)=06:59:31.588, Dec.(J2000)=-04:05:27.86, i.e. it is at $\Delta \alpha$=-15\,mas, $\Delta \delta$=-20\,mas from the FUor,
that is $\Delta r$=25\,mas (or 11.25\,au), with a P.A.=233\fdg1. A rough estimate of the Pa$\beta$ and Br$\gamma$ line fluxes gives $\sim$7$\times$10$^{-15}$ and 2$\times$10$^{-15}$\,erg\,s$^{-1}$\,cm$^{-2}$, respectively.     
The detection in both continuum-subtracted line images, the fact that these features are fitted with Gaussian profiles, and the presence of continuum emission in the PSF-subtracted $J$ and $K$ band images strongly suggest that the `bump' is an additional closer young companion, which might have triggered the burst~\citep[see e.g.,][]{reipurth04,reipurth14}. Moreover, in both $J$ and $K$ bands, the `bump' appears brighter than 2MASS\,J06593158-0405277B, hinting that,
if it is a YSO, it might be more massive than the latter. However, as the `bump' is not spatially resolved in the continuum, other scenarios cannot be completely ruled out for the continuum emission, such as an uneven disk illumination~\citep[see, e.g.,][]{dullemond10}.

Finally, the lower panels of Fig.~\ref{fig2} show the FUor flux calibrated spectrum ($m_J$=8.63$\pm$0.05\,mag and $m_K$=7.45$\pm$0.04\,mag), which only exhibits molecular and atomic features in absorption, as usually observed in 
FUor spectra, namely H$_2$O, CO, \ion{H}{1} (Pa$\beta$ and Br$\gamma$), \ion{Na}{1}, \ion{Ca}{2},
\ion{K}{1}, \ion{Mg}{1}, and \ion{Fe}{1}. From these features we infer a spectral type between K9 and M1, typical of FUor NIR spectra~\citep[see, e.g.,][]{audard}. Notably, this spectrum closely resembles that of 
\objectname{HBC 722}, a well studied FUor~\citep[see,][]{miller}.

\section{DISCUSSION} \label{discussion}

The detection of binaries or higher order systems in FUors is potentially of great importance, as interaction with
a companion could be a triggering mechanism, although, in principle, there might be more than one. For example, a couple of FUors (V1057\,Cyg and V1515\,Cyg) are probably single systems~\citep[see][and references therein]{audard}, thus, in this case, internal instabilities seem to be the most straightforward process to consider. 
On the other hand, a large number of FUors and FUor candidates are binaries,
and \citet{bonnell92} first proposed that FUor events may be triggered by the passage of a close companion. Later on, \citet{reipurth04} suggested that these objects might be newborn binaries, which arise from the breaking of unstable triple systems~\citep[see also][]{reipurth14}.
As a consequence of the dynamical interaction, one object is ejected, typically the lowest mass component, whereas the remaining two objects spiral in toward each other perturbing their disks. The bursts would be then triggered by the close passage ($\sim$10\,au) of the nearest companion.

This scheme is a plausible explanation for what we are observing in 2MASS J06593158-0405277. The central object is a FUor, whose burst might have been triggered by a close companion (the `bump') 
located at $\sim$11\,au, and there is a more distant ($\sim$100\,au) and less massive object, which could be the third (ejected?) companion of a non-hierarchical 
system~\citep[e.g.,][]{reipurth14}. Notably, the present configuration is precisely what would be expected from the dynamical evolution of a
newborn triple system, where the least massive of three bodies is kept away from the two other bodies except for brief periastron
passages, which eventually lead to the system dissolution~\citep[see, e.g.,][]{reipurth01,reipurth10}. At an age of $\sim$1\,Myr the observed system is statistically unlikely to be stable, 
and it will eventually eject the low-mass component after a periastron passage.
This could be the first time that such scenario has been observed. Indeed 
companions in other FUors have been detected~\citep[e.g., FU Ori, Z\,CMa, L1551\,IRS\,5, RNO 1B/C, AR 6A/B; see e.g.,][]{reipurth04,audard}, 
but their relatively wide separation (from several tens up to hundreds of au) suggests that these sources might be outlying companions. 

In the case of 2MASS J06593158-0405277, the fact that both the `bump' and source `B' are young objects excludes the possibility that they might be background or foreground stars. Instead it is much more likely that 
they belong to the same star forming region, which is quite small and isolated, being composed of a few T\,Tauri stars~\citep[][]{kospal}. Their proximity to the FUor and the fact that the 
observed radial velocities of the HI lines are very similar in the three objects (average $v_{r}$$\sim$7$\pm$7\,km\,s$^{-1}$ with respect to the local standard of rest -LSR-
and not corrected for the molecular cloud velocity) suggests that 
these sources might belong to the same system.
Provided that the objects are physically linked, 2MASS\,J06593158-0405277 would then be one of the very few triple systems ever observed in FUors. That said, further high-angular resolution observations are 
required to confirm the nature of the `bump' and the multiplicity of the 2MASS J06593158-0405277 system.

\acknowledgments{
A.C.G., R.G.L., and T.P.R. were supported by Science Foundation Ireland, grant 13/ERC/I2907.
S.K. acknowledges support from an STFC Ernest Rutherford Fellowship (ST/J004030/1), Ernest Rutherford
Grant (ST/K003445/1), and Marie Curie CIG grant (SH-06192).
}
{\it Facilities:} \facility{ESO/VLT}.

\end{document}